\documentclass[11pt,twoside]{article}

\usepackage{asp2006}
\usepackage{epsf}
\usepackage{epsfig} 
\usepackage{lscape}

\begin{document}
\title{Age, metallicity and $\alpha$-elements abundance in stellar population}
\author{M. Koleva$^{1,2}$, R. Gupta$^3$, Ph. Prugniel$^{1,4}$, and H. Singh$^3$}
\affil{
1 Universit\'{e} de Lyon 1, CRAL, Observatoire de Lyon, CNRS, UMR5574 ; ENS de Lyon, France\\
2 Department of astronomy, University of Sofia, Bulgaria \\
3 IUCAA, Pune, India \\
4 GEPI, Observatoire de Paris, France \\
}    

\begin{abstract} 

We built modelled spectra of stellar population at high resolution and with 
variable $\alpha$-elements enhancements.
Analysing spectra of Galactic globular clusters we show that it is possible 
to derive reliably and efficiently [Mg/Fe] using spectra integrated along the 
line-of-sight.
These detailed measurements open perspectives for investigating the enrichment 
process on galaxies and star clusters.
\end{abstract}


{\noindent \bf Flux recalibrated CFLIB library.} The CFLIB stellar library 
(or IndoUS; \citealt{indous}) contains spectra of 1273
 stars observed in the wavelength range 3460-9464 $\AA$ at a spectral 
resolution 
FWHM=1~$\AA$ (R$\approx$5000). The selection of the stars provides a good 
coverage of the 
T$_{eff}$, $\log g$, [Fe/H] space.
In the initial release, the spectra were flux calibrated using the templates 
of \citet{pick98}. This calibration was inaccurate for two main reasons: 
(1) the metallicity dependency of the spectral energy distribution (SED)
was neglected, and (2)
it relied on the precision of the atmospheric parameters.
To generate high quality stellar population models, 
we corrected the flux calibration using the MILES library (\citealt{miles})
that has a lower spectral resolution, but has almost the same spectral coverage
and an accurate flux calibration.
The re-calibrated library will be publicly available in the near future.

%

{\noindent \bf Modelled spectra of stellar population with $\alpha$-enhancement.}To build population models with parametric $\alpha$-elements we followed the recipe of 
\citet{php07}: The empirical library, supposed to have the $\alpha/$Fe pattern of the solar
neighbourhood, is differentially corrected for the effect of variable 
abundance using the \citet{coelho} library. Two semi-empirical grids are produced with
$[Mg/Fe]=0$ and $+0.4$. Intermediate values are linearly interpolated as a
 function of the mass ratio of Mg to Fe.

Single stellar population (SSP) models are built using Pegase.HR (\citealt{phr})
 and the solar-scaled isochrones of Padova group. The drawback of not using enhanced
 isochrones (\citealt{sal00}) may essentially affect the ages
 (the enhanced isochrones are bluer). 

To determine the age, [Fe/H] and [Mg/Fe] of a population from its spectrum, we use the full
 spectrum fitting method discussed in \citet{kol06}. 
It is implemented as a simple $\chi^2$ minimisation where the free physical parameters 
of the model are those of  the population mix, plus the internal kinematics. 
In addition to the physical free parameters, the model contains a multiplicative 
polynomial making the process insensitive to the shape of the continuum, just 
like the classical Lick indices. This procedure has been found to be reliable and
 to give a three times better precision than Lick indices due to the optimised 
usage of the information. We performed Monte-carlo simulations and we have seen that
 there is no degeneracy
 between Age-[Mg/Fe] or [Fe/H]-[Mg/Fe]. The absence of degeneracy between [Mg/Fe]
 and age allows to grant some confidence to the retrieved values of [Mg/Fe], 
since the uncertainties on the evolutionary tracks will mostly bias the age
and not the other parameters.

 \setcounter{figure}{0}
 \begin{figure}[!h]
 \plottwo{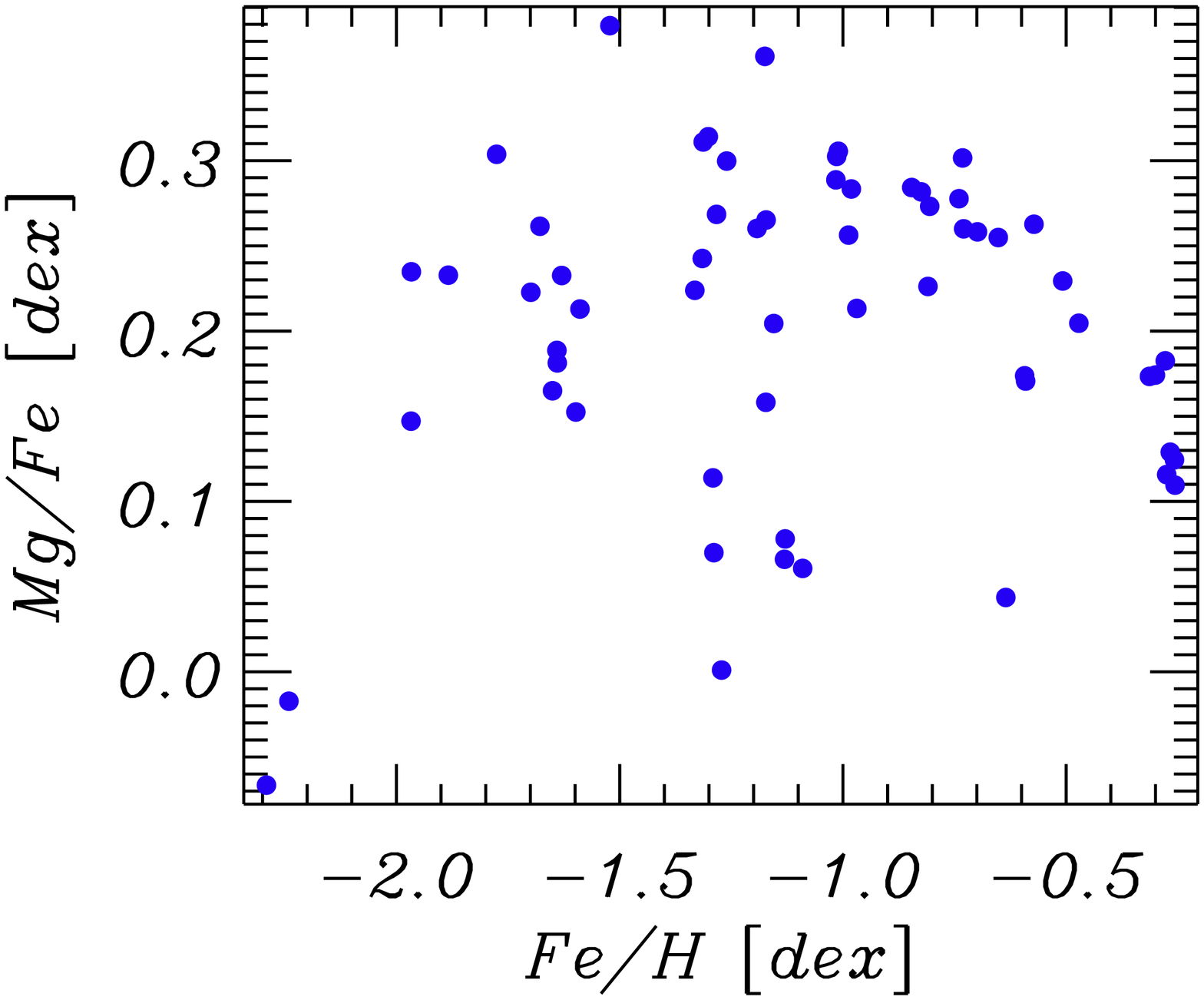}{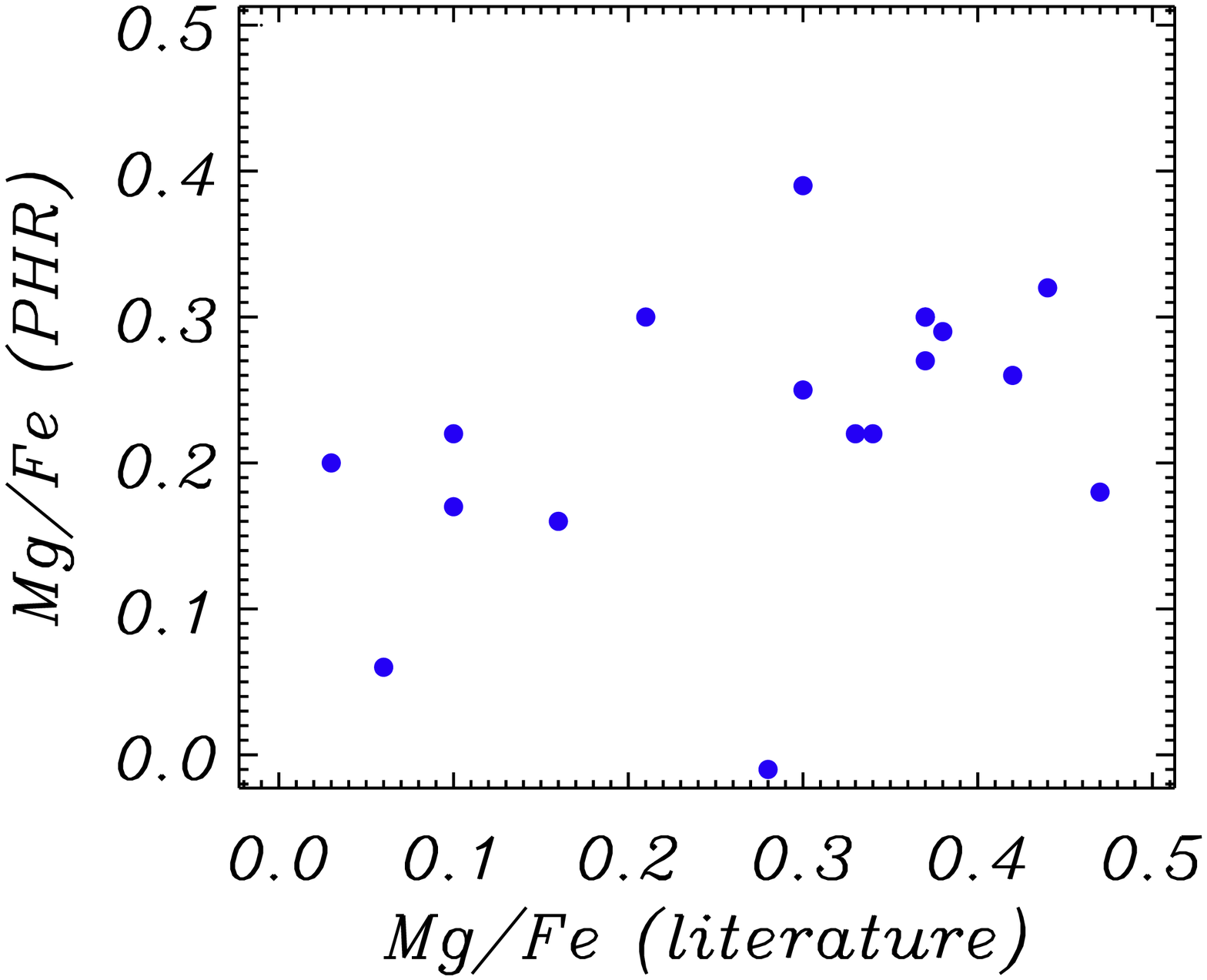}
 \caption{{\itshape Left:\/} Distribution of the [Mg$/$Fe] as a function of [Fe$/$H] for GGCs.
 {\itshape Right:\/}  Comparison between our values and the one from the literature.
 }
 \end{figure}

{\noindent \bf Application to globular clusters.}We analysed the spectra of Galactic globular clusters from \citet{ggc}.
 The distribution of [Mg/Fe] is shown on Fig. 1 with a comparison with
 the compilation of spectroscopic abundances (\citealt{pri05}, \citealt{tan07}).
There is a satisfactory agreement between the enhancement measured from stellar
 spectroscopy and our measurements from line-of-sight integrated spectra.
The measurement of [Mg/Fe] from integrated spectra using full spectrum 
fitting is reliable and more precise than with Lick indices.




\end{document}